# Control of ferromagnetism of Vanadium Oxide thin films by oxidation states

*Kwonjin Park, Jaeyong Cho, Soobeom Lee, Jaehun Cho, Jae-Hyun Ha, Jinyong Jung, Dongryul Kim, Won-Chang Choi, Jung-Il Hong, and Chun-Yeol You[*]*

K.Park, J.Cho, J.-H.Ha, J.Jung, D.Kim, W.-C.Choi, J.-I.Hong, C.-Y.You.

Department of Physics & Chemistry, Daegu Gyeongbuk Institute of Science & Technology (DGIST), Daegu 42988, Republic of Korea

E-mail addresses: cyyou@dgist.ac.kr (Chun-Yeol You)

S.Lee

Department of Electrical and Computer Engineering, Shinshu University, Nagano 380-0928, Japan

J.Cho

Division of Nanotechnology, Daegu Gyeongbuk Institute of Science & Technology (DGIST), Daegu 42988, Republic of Korea

**Abstract**

Vanadium oxide ($VO_x$) is a material of significant interest due to its metal-insulator transition (MIT) properties as well as its diverse stable antiferromagnetism depending on the valence states of V and O with distinct MIT transitions and Néel temperatures. Although several studies reported the ferromagnetism in the $VO_x$, it was mostly associated with impurities or defects, and pure $VO_x$ has rarely been reported as ferromagnetic. Our research presents clear evidence of ferromagnetism in the $VO_x$ thin films, exhibiting a saturation magnetization of approximately 14 kA/m at 300 K. We fabricated 20-nm thick $VO_x$ thin films via reactive sputtering from a metallic vanadium target in various oxygen atmosphere. The oxidation states of ferromagnetic $VO_x$ films show an ill-defined stoichiometry of $V_2O_{3+p}$, where $p$ = 0.05, 0.23, 0.49, with predominantly disordered microstructures. Ferromagnetic nature of these $VO_x$ films is confirmed through a strong antiferromagnetic exchange coupling with the neighboring ferromagnetic layer in the $VO_x$/Co bilayers, in which the spin configurations of Co layer is influenced strongly due to the additional anisotropy introduced by $VO_x$ layer. The present study highlights the potential of $VO_x$ as an emerging functional magnetic material with tunability by oxidation states for modern spintronic applications.

**Introduction**

The electrical properties of a series of vanadium oxides ($VO_x$) represent one of the most extensively studied topics since the discovery of the metal-insulator transition (MIT).[1–3] It has been observed that various valence states of $VO_x$ exhibit distinct MIT temperatures, which systematically depend on the valence numbers of the oxides. These transitions are invariably accompanied by crystal phase changes and significant electron-electron correlations. Numerous comprehensive review articles have addressed the MIT and electrical properties of $VO_x$. Furthermore, it is well-established that $VO_x$ exhibits antiferromagnetism at low temperatures when $x < 2.0$. Typically, the Néel temperature decreases with increasing $x$ (for $x > 1.5$), transitioning to paramagnetism at $x = 2.0$.[2] The absence of ferromagnetic phases in $VO_x$ has led to a perception that the study of its magnetic properties is unremarkable and has garnered limited attention, as the magnetic applications of antiferromagnetic and paramagnetic $VO_x$ are constrained.

Several reports have documented the occurrence of ferromagnetic phases in $VO_x$.[4–8] For instance, room-temperature ferromagnetism has been observed in I- or Li-doped nanotube structures. These studies suggest that the origin of ferromagnetism was attributed to the frustrated vanadium spins in self-assembled nanotubes induced by electron or hole doping [4] Another approach involves the introduction of defects with unpaired electrons ($V^{4+}$) by exploiting the twin boundaries of epitaxially grown $VO_2$ structures without chemical reactions, which resulted in a saturation magnetization of approximately 18 kA/m at room temperature and a high Curie temperature (~ 500 K).[5] Moreover, in nanowire structures utilizing $VO_2$, annealing facilitates the creation of oxygen vacancies, leading to the formation of superparamagnetic chains of $V^{3+}$-$V^{4+}$ dimers from singlet $V^{4+}$-$V^{4+}$ dimers, thereby resulting in weak ferromagnetism.[6] In addition to VO$_2$, research has also been directed towards V$_2$O$_5$, a polycrystalline material, where the generation of oxygen vacancies induces changes in structural, electrical, and magnetic properties.[7,9,10] This study reported the coexistence of antiferromagnetism and ferromagnetism, contingent upon the degree of vacancy, leading to variations in the dominant magnetic phase as elucidated through density functional theory. Collectively, these prior investigations strongly support the potential for ferromagnetic VO$_x$ phases through structural variations or alterations in oxygen vacancy content.

In this study, we present clear experimental evidence for the existence of a ferromagnetic phase in 20 nm thick VO$_x$ thin films. At room temperature, we observed a magnetization of 14 kA/m and a magnetic susceptibility ($\chi$~10). Although the magnetization and susceptibility values are lower than those typically observed in strong ferromagnetic materials, they are significantly greater than those found in typical paramagnetic and/or antiferromagnetic phases ($|\chi| < 10^{-5}$), with the Curie temperature exceeding room temperature. The 20 nm VO$_x$ thin films were fabricated using a reactive sputtering system with a metallic vanadium target on thermally oxidized silicon wafers. During the sputtering process, we meticulously controlled the O$_2$ gas flow to achieve non-stoichiometric VO$_x$ phases. The resulting samples do not belong to the well-known specifically V$_n$O$_{2n-1}$ (Magnéli phases), V$_n$O$_{2n}$, nor V$_n$O$_{2n+1}$ phases,[11] but rather consist of a mixture of various valence states of vanadium. Our analysis reveals the presence of vanadium ions in the V$^{3+}$ and/or V$^{4+}$ states, with their relative concentrations dependent on the composition and O$_2$ gas flow rate. We speculate that the unpaired V ions are the source of the observed ferromagnetism. The observed ferromagnetism is somewhat surprising because V$_n$O$_{2n-1}$ are antiferromagnet for $n$ = 2-8 and VO$_2$ (corresponding to $n\rightarrow \infty$) is paramagnet,[12,13] while our VO$_x$ samples ($x$= 1.53, 1.65, 1.72) are close to the V$_2$O$_3$ antiferromagnetic phase when we consider the composition only. According to the *ab initio* calculations using spin DFT (density function theory),[14] magnetic moment of V atom is 1~2 μ$_B$ (Bohr magnetron), but they form antiferromagnetic phase so that net moment of VO$_x$ is almost zero. But, the lack of oxygen atom, oxygen vacancies, induce a ferromagnetic phase. It must be noted that the magnetization of our Sample 1 is only 1~2 % of typical ferromagnet whose atomic moment is ~ 2 μ$_B$, and we will explain it later.

Additionally, we discovered significant exchange coupling between the VO$_x$ layers and an adjacent ultra-thin Co layer, which serves as one of the signatures and possible functionality of the ferromagnetic VO$_x$ phase. A detailed examination of the exchange coupling in VO$_x$/Co/Pt trilayer structures indicates that the spin configurations of the Co layer are notably influenced by the VO$_x$ layers. This suggests that, despite the relatively low net magnetization of the VO$_x$ layers, the atomic moments are on the order of $\mu_B$ (Bohr magneton), countering previous predictions by Xiao.[7,9] Consequently, the ferromagnetic VO$_x$ phase has the potential to function as a novel magnetic material, synergizing with its unique electrical properties for emerging applications. We believe that our findings broaden the scope of VO$_x$ usage in next-generation multifunctional devices.

1. **Results and discussion**

Figure 1 illustrates the magnetization hysteresis curves (a, b) and the susceptibility ($\chi = \partial M/\partial H$) (c, d) of various VO$_x$ films with a thickness of 20 nm at temperatures of 10 K and 300 K, respectively. Samples 1, 2, and 3 were fabricated using different oxygen flow rates of 2.5, 3.5, and 4.5 sccm during the magnetron sputtering process (refer to Section 4: Experimental for detailed sample preparation methods). Detailed information on Samples 1, 2, and 3 is provided in Table I, alongside comparative data for $V_2O_3$ and $VO_2$. It is important to note that $V_2O_3$ and $VO_2$ samples were also prepared in the same chamber but under different oxygen gas flow conditions. The susceptibility $\chi \sim 0$ for $V_2O_3$ and $VO_2$ at both 10 K and 300 K suggests that these samples exhibit antiferromagnetic and/or paramagnetic phases.

Focusing on Samples 1, 2, and 3, particularly Sample 1, which was prepared under the lowest oxidation conditions, a distinct hysteresis loop and $\chi \sim 10$ at both 10 and 300 K indicate ferromagnetism, as shown in Figure 1a-d. In these measurements, the paramagnetic signal from the 550 μm thick $Si/SiO_x$ substrate is significant due to its relatively large volume, necessitating careful analysis to accurately obtain the magnetization hysteresis loops. To mitigate potential errors, the dimensionless susceptibility $\chi = \partial M/\partial H$ was calculated, where the paramagnetic contribution from the substrate introduces an offset of approximately $10^{-5}$. While $\chi \sim 10$ is small compared to typical ferromagnetic materials, it is significantly larger than the typical values for paramagnetic and/or antiferromagnetic phases ($|\chi| < 10^{-5}$). The observed magnetization of 14 kA/m is also considerably lower than that of typical ferromagnets (~1000 kA/m). We will later discuss the physical reasons of the 1-2% magnetization of $VO_x$. Initially, such a low magnetization value may suggest weak ferromagnetism or altermagnet.[15,16] However, a rigorous definition of weak ferromagnetism and altermagnet require a well-defined crystal structure, which is not applicable in this context. Additionally, weak ferromagnetism originates from a slight non-collinear spin alignment of isomorphous atoms in antiferromagnetic materials, resulting in a finite tilting angle and a non-zero magnetic moment. The altermagnet necessitates specific crystal symmetries to exhibit non-degenerate k-space dependent spin splitting, which is not applicable to our cases[17,18]. We hypothesize that the observed ferromagnetic properties are due to small volumes of ferromagnetic phases within an antiferromagnetic phase matrix. This ferromagnet-antiferromagnet composite model will be discussed in detail later. Consequently, we refrain from labeling our findings as "weak ferromagnetism" despite the low magnetization values. Furthermore, the Curie temperature is evidently higher than room temperature, indicating that our ferromagnetic $VO_x$ can be utilized in room temperature magnetic devices.

The magnetization and susceptibility of Samples 2 and 3, prepared under more oxidizing conditions, are smaller than those of Sample 1, yet they still exhibit ferromagnetic phases. Sample 1, with the lowest oxidation condition, exhibits the highest susceptibility, which decreases gradually with increasing oxidation. Samples 1, 2, and 3 were intentionally prepared under non-stoichiometric conditions to create mixtures of $V^{3+}$ and/or $V^{4+}$ ions, with the ratio of each ion controlled by the oxygen gas flow rates. Previous studies suggest that unpaired electrons ($V^{4+}$) and/or oxygen vacancies may serve as the origin of ferromagnetism. Therefore, we speculate that the observed ferromagnetism is closely related to the valence state of oxygen.

To elucidate the physical origin of the ferromagnetism observed in the samples, it is crucial to quantify the relative amounts of $V^{3+}$ and $V^{4+}$ ions in Samples 1, 2, and 3. We employed electron energy loss spectroscopy (EELS) to determine these contributions. Figure 2a presents the energy loss spectra of single-crystal $V_2O_3$ and $VO_2$ (Figure S1, Supporting information), as well as the changes in oxidation states of $VO_x$ (Sample 1, 2, and 3) using TEM-EELS. As illustrated, the energy loss peaks of the V-$L_3$ edge shift to higher energies with increasing oxidation, transitioning from $V_2O_3$ to $VO_2$. Additionally, in the oxygen K-edge region, the broadening of the spectrum changes progressively with the oxidation state, indicating that the oxidation levels of $VO_x$ in Samples 1, 2, and 3 lie between those of $V_2O_3$ and $VO_2$.

To quantify the amounts of $V^{3+}$ and $V^{4+}$ ions in Samples 1, 2, and 3, we utilized three distinct methods based on EELS spectra. The first method involves comparing the areas under the energy loss spectra for $V^{3+}$ and $V^{4+}$ ions. The contribution of $V^{3+}$ is denoted by $\alpha L_3(V_2O_3)$, while that of $V^{4+}$ is denoted by $\beta L_3(VO_2)$. As oxidation increases, $\alpha$ decreases and $\beta$ increases linearly. This method allows us to determine the relative contributions of $V^{3+}$ and $V^{4+}$ in Samples 1, 2, and 3. The second method involves comparing the intensity ratio of the $L_3$ to $L_2$ edges, $I(L_3)/I(L_2)$, which is also indicative of the oxidation states of $VO_x$. Finally, we examined the linear variation in the peak position of the $L_3$ edge in the energy loss spectra, which is known to depend linearly on the oxidation state.[19] All three EELS analysis methods[20,21] yielded consistent compositions of $V^{3+}$ and $V^{4+}$ ions for Samples 1, 2, and 3, as summarized in Table I. This consistency provides evidence that $V(III, IV)O_x$ exists as a mixed state of $V^{3+}$ and $V^{4+}$ ions. Further details of the analysis methods are provided in Figure S2 (Supporting information).

High-resolution transmission electron microscopy (HR-TEM) was employed to investigate the microstructural characteristics of the $VO_x$ samples. Figures 3a-c present the HR-TEM images, where the crystal lattice is scarcely discernible, suggesting a predominantly amorphous phase. To facilitate further analysis, Fast Fourier Transformation (FFT) was carried out as shown in Figure 3d displaying an amorphous nature in $VO_x$ for Sample 1A. Additionally, Figures 3e and 3f display the results for Sample 2A and 3A, respectively, indicating that with increasing oxidation levels, a weak crystal structure begins to emerge gradually. To substantiate these observations, line profiles were extracted by integrating the diffraction patterns corresponding to the main ring, as shown in Figures 3g-i. These profiles were obtained by fitting the intensity variations, excluding background noise, to provide qualitative insights based on the broadness of the peaks. For comparison, single-crystal TEM results of $V_2O_3$ and $VO_2$ are presented in Figure S2, demonstrating clearly ordered crystalline structures. Although deriving quantitative information from the intensity alone is challenging, the breadth of the peak offers meaningful insight on the atomic structures. In Sample 1A, which was prepared under the lowest oxidation conditions, the material exhibits a nearly amorphous state. In contrast, Samples 2A and 3A show diffuse rings in the diffraction patterns with diameters corresponding approximately to the consistent interplanar distances and suggest indicating the presence of weak short-range ordering toward the oxide lattices. Furthermore, in Sample 3A, points corresponding to a larger distance are observed. These findings suggest that a lattice ordering begins gradually as $V^{3+}$ becomes more prevalent than $V^{4+}$ Due to the changes in oxidation levels during the deposition.

Typically, crystalline $V_2O_3$ exhibits an insulating phase with antiferromagnetic (AFM) order. As oxidation increases, the coupling strength between $V^{3+}$ ions diminish, and additional $V^{4+}$ ions are incorporated. These additional isolated $V^{4+}$ ions contribute to the emergence of a ferromagnetic phase.[6] The relatively small magnetization observed in Samples 2A and 3A can be explained by the following: with increased oxidation, the number of $V^{4+}$ ions rise. A higher density of $V^{4+}$ ions lead to antiferromagnetic coupling between $V^{4+}$ ions, which gradually reduces the volume of the ferromagnetic phase.

In ordered to confirm that the measured ferromagnetism is from the V oxide phase itself, rather than the possible presence of magnetic minor phase in the samples, along with the consideration of extending the functionality of the $VO_x$ phase for potential applications, we investigated the exchange coupling of the $VO_x$ with a conventional ferromagnetic layer. A series of $VO_x$ (20 nm)/Co (1.4 nm)/Pt (5 nm) trilayer samples (Samples 1A, 2A, 3A) were prepared, where the bottom VO$_x$ layers are nominally identical to those in Samples 1, 2, and 3. The top 5 nm Pt layer is included to promote perpendicular magnetic anisotropy. Figure 4a displays the measured magnetic moment as the temperature is increased from 10 K without applying an external field. During the cooling process to 10 K, an in-plane magnetic field (1 T) is applied to saturate the samples. At 10 K, both Sample 1 and Sample 1A exhibit a positive magnetic moment (labeled "$M_L$" in Figure 4a), as shown in the hysteresis loop in Figure 4b. However, as the temperature increases, the magnetic moment of VO$_x$ single layer decreases but retains a positive value, while Sample 1A shows a switching from positive to negative magnetic moment at 144 K which is postulated to be due to a switch of Co layer magnetization from a positive to a negative spin arrangement at room temperature (labeled "$M_R$" in Figure 4a), as depicted in the hysteresis loop in Figure 4c. This behavior suggests strong antiferromagnetic coupling between the VO$_x$ and the ferromagnetic (FM) layers. It is also noteworthy that the magnitude of magnetic moment for Sample 1A is greater at room temperature than at low temperature. This can be explained by the fact that at low temperature ("$M_L$"), the value is not saturated and is nearly zero field value, as shown in Figure 4b. In contrast, the relatively large absolute value of the magnetic moment at room temperature ("$M_R$") is close to the saturation value, as indicated in Figure 4c.

To confirm the strong antiferromagnetic exchange coupling between VO$_x$ and Co layers, hysteresis loops were examined at 10 and 300 K for Sample 1A, as displayed in Figures 4b and 4c. Typically, the magnetization of Co is known to be around 1400 kA/m, yet it exhibits a lower value of 800 kA/m in the trilayer, indicating the presence of a finite ferromagnetic phase with misaligned spin arrangement. A normal hysteresis loop is observed at low temperatures (10 K), while an inverted loop is observed at room temperature (300 K), indicating that switching occurs before the field direction changes. Inverted loops are known to occur due to various causes, particularly attributed to antiferromagnetic coupling. It is postulated that a competition between antiferromagnetic exchange bias and ferromagnetic anisotropy energy[22,23] exist in the Co layer due to the significant antiferromagnetic exchange coupling between the VO$_x$ and the FM Co layer..

Based on Figures 4a, b, and c, it can be inferred that the VO$_x$ layer has a substantial magnetic moment which is antiferromagnetically coupled with Co layer. Similar trends were observed for Samples 2A and 3A with different coupling strength with Co layer. It is important to mention that if Samples 1, 2, and 3 were weak ferromagnets with a fully antiferromagnetic phase and a non-zero magnetic moment due to a small tilting angle, the observed data could not be explained. Because the Néel temperature of VO$_x$ phases is nominally far below 200 K,[2] the VO$_x$ layer is expected to be paramagnetic at 300 K. Furthermore, the significant exchange coupling implies that the atomic moment is at least on the order of $\mu_B$; otherwise, the coupling strength would be orders of magnitude smaller. The ~$\mu_B$ atomic moment seems contradictory to the small magnetization of the VO$_x$ layer, but this contradiction can be resolved by a partial volume ferromagnetic phase model, which will be discussed later.

To elucidate the influence of the VO$_x$ layer on the spin configuration of Co layers in Samples 1A, 2A, and 3A, azimuthal angle-dependent Hall effect measurements were conducted. These measurements help discern how the VO$_x$ layer affects the magnetic properties of the Co layer, particularly in terms of its magnetization direction. When an applied magnetic field is sufficiently strong to saturate the magnetization in the direction of the field, the magnetization vector $\vec{M}$ aligns parallel to the field. The magnetization direction vector can be expressed as $\hat{m} = (\sin\theta_M \cos\varphi_M, \sin\theta_M \sin\varphi_M, \cos\theta_M)$ where $\theta_M$ is the polar angle and $\varphi_M$ is the azimuthal angle of the magnetization vector in the typical spherical coordinate system. The effect is referred as planar Hall effect (PHE).[24,25] Or anomalous Hall effect (AHE) depending on the direction of magnetization with respect to the film plane corresponding to $\theta_M = \frac{\pi}{2}$ or 0, respectively. The total Hall resistance $R_{xy}$ would be represented as the sum of both AHE and PHE contribution as belows:[26]

$$R_{xy} = \frac{1}{2}\Delta R_{AHE} \cos\theta_M + \frac{1}{2}\Delta R_{PHE}\sin^2\theta_M \sin 2\varphi_M \qquad (1)$$

The current measurement is taken for the purpose of investigating the exchange interaction between $VO_x$ and Co layers. Given the anticipated weak interaction, we chose to perform measurements with a slight tilt ($\theta_t$) rather than aligning completely with either the *xy* plane or the *xz* plane (as depicted in Figure 5a). This approach allows more sensitive detection of subtle changes arising from the exchange interaction that might be obscured in a strictly planar or perpendicular configuration. The detailed measurement schemes, including the specific setup and rationale behind using a tilted configuration, are elaborated in Note S1 (Supporting information). This note provides additional context on how the tilt angle ($\theta_t$) was selected and the ways in which it enhances the sensitivity of our measurements to the exchange interaction.

In this experiment, we fixed the tilt angle $\theta_t = -10°$ and applied an external rotating in-plane magnetic field of 0.1 T and measured $R_{xy}$ as a function of azimuthal angle $\varphi_H$, as shown in Figure 5a. Because we employed tilted geometry, the $R_{xy}$ exhibits $\sin\varphi_H$ and/or $\sin 2\varphi_H$ dependences corresponding to the out-of-plane and in-plane components of the magnetization (corresponding to the AHE and PHE), respectively. It is noted that the perpendicular magnetic anisotropy (PMA) is strong enough at 10 K, hence the 0.1 T in-plane field is insufficient to fully align the magnetization along the external field in the polar direction, meaning that $\theta_M \neq$ $\frac{\pi}{2} - \theta_t$. However, it is strong enough to align the magnetization in the azimuthal direction ($\varphi_M = \varphi_H$) for the control sample (Co/Pt without the VO$_x$ layer), as shown in Figure 5b-I. Therefore, without precisely determining $\theta_M$, $R_{xy}$ shows a $\sin \varphi_H$ dependence, indicating that AHE is dominant, and the PHE is negligible, suggesting a small $\theta_M$ under an in-plane field of 0.1 T in our experimental geometry. When a stronger in-plane external field is applied, the $\sin \varphi_H$ (AHE contribution) diminishes, and the $\sin 2\varphi_H$ (PHE contribution) becomes dominant, as shown in Figure 3S (Supporting information).

For Samples 1A (red), 2A (blue), and 3A (olive), the $\varphi_H$ dependent $R_{xy}$ is depicted in Figures 5b II-IV, respectively. Compared to the control sample (Figure 5b-I), these samples exhibit a $\sin \varphi_H$ dependence with a finite phase shift. These non-zero phase shifts imply that the assumption $\varphi_M = \varphi_H$ is no longer valid, providing further evidence of significant exchange coupling between the Co and VO$_x$ layers. The phase shifts are more clearly illustrated in polar coordinate plots in Figure 5c. The phase shift is most pronounced in Sample 1A, which has the largest magnetic moment, and nearly disappears for Sample 3A, which has the smallest magnetic moment. These non-zero phase shifts in $R_{xy}(\varphi_H)$ indicate additional in-plane anisotropy induced by the VO$_x$ layer. The magnetic VO$_x$ layer introduces an in-plane anisotropy that can be controlled by varying the oxidation conditions, revealing a new functionality of the magnetic VO$_x$ layer that has not been previously reported. This finding highlights the potential for tailoring the magnetic properties of these materials for specific applications by adjusting the oxidation process.

In Figures 2 and 3, it is indicated that the VO$_x$ layer contains a mixture of V$^{3+}$ and V$^{4+}$, and when V$^{3+}$ is predominant and V$^{4+}$ is less prevalent, there is a possibility of ferromagnetism emerging in Samples 1, 2, and 3. This suggests that the ferromagnetic phase might be present as randomly distributed grains within the VO$_x$ layer. Based on this observation, we propose a "partial volume fraction ferromagnetic phase model," where approximately 10% of the volume fraction is considered to be in the ferromagnetic phase, while the remaining portion is in the antiferromagnetic or paramagnetic phase. This model provides a plausible explanation for the experimental results we've observed. It's important to note that while our model offers a reasonable interpretation of the findings, it is not the sole explanation. The complexity of the system means that other models could potentially account for the observed behavior as well. To gain a deeper understanding of the ferromagnetic behavior in VO$_x$, more direct evidence, such as detailed measurements of the atomic moments and spin configurations within the VO$_x$ layer, would be necessary. Unfortunately, such investigations are beyond the scope of our current study. Further research using advanced characterization techniques, such as X-ray magnetic circular dichroism (XMCD) or neutron scattering, could provide more insights into the magnetic properties and confirm the validity of the proposed model.

To confirm the "partial volume fraction ferromagnetic phase model", we carried out micromagnetic simulations with MuMax3.[27] Details of simulations are described in 4. Experimental section. We simulated single VO$_x$ layer and VO$_x$/FM bilayer. The VO$_x$ are divided into 200 grains with randomly distributed anisotropy energy and easy axis direction to mimic the amorphous nature of VO$_x$ layer. And we set ferromagnetic exchange coupling for only 2% of grains (4 grains), and remaining grains have antiferromagnetic exchange coupling. We used random inter-grain exchange coupling energy. We assumed that the magnetization of VO$_x$ layer is $M_{s,V}$ (= $1.0 \times 10^6$ A/m). It is comparable value with the typical ferromagnet materials, and much larger than our experimental results. We will explain the reason of such large value of $M_{s,V}$. And FM layer has typical ferromagnetic Co properties with PMA (=$0.7 \times 10^6$ J/m$^3$).

Figure 6a presents the micromagnetic simulation hysteresis loop of the *x*-component magnetization as a function of external in-plane magnetic fields (ranging from -5 T to +5 T) along the *x*-axis for a single VO$_x$ layer. The plot focuses on the region from -1 T to +1 T to highlight the details in the low-field region. We adjusted the material parameters of VO$_x$ to replicate the experimentally observed hysteresis loop (refer to Fig. 1a). It is important to note that the simulated magnetization is 15 kA/m, which represents only 1.5% of the input value for VO$_x$ ($M_s$ = 1.0 × 10$^6$ A/m). We set the $M_s$ value of VO$_x$ to that of typical ferromagnetic materials to reflect the reported atomic moment of the V$^{4+}$ ion (~1 $\mu_B$). Therefore, the result of 15 kA/m is not unexpected, given that only 2% of the volume is attributed to the ferromagnetic phase, while the remaining portion consists of the antiferromagnetic phase. The spins in the antiferromagnetic phase effectively cancel each other out, do not contributing net magnetization. Furthermore, the interfacial spins between the ferromagnetic and antiferromagnetic phases are also exchange coupled, so that the interface spins of the ferromagnetic phase cannot contribute to the net moment in small field region. Consequently, the observed 1.5% of $M_s$ for VO$_x$ is consistent within this simulation framework, implying that the actual volume fraction of the ferromagnetic phase in our sample is likely order of 1 %. Our partial volume fraction ferromagnetic phase model provides an explanation for the experimental observations of the low magnetization (approximately 1% of that of a typical ferromagnet) corresponding to the atomic moment of the V$^{4+}$ ion (~ 1 $\mu_B$). At first glance, this may appear contradictory; however, the partial volume fraction ferromagnetic phase model effectively resolves this discrepancy.

In Figure 6b, the spin configuration in the VO$_x$ single layer was visually demonstrated using MuView[28] at $H_x$ = 0.4 T. In this illustration, the small portion of the ferromagnetic phase surrounded by the antiferromagnetic phases are clearly shown. More details of the spin configuration of ferromagnetic and antiferromagnetic phases are shown in the insets. Only 2 % of volume fractions are ferromagnetic phases in our simulations, which are easily saturated but pinned with the surrounding antiferromagnetic phases. The pinning caused the slanted loop shapes in Fig. 6a. At 0.4 T of external magnetic field, it is only partially saturated due to the antiferromagnetic phase. This behavior qualitatively aligns well with the experimental results.

Next, we try to reproduce the experiments of the phase shift in $R_{xy}(\varphi_H)$ of $VO_x$/Co/Pt trilayer samples in Figure 5b I-IV, which is clear evidence of the non-zero magnetic coupling between $VO_x$ and Co layers. We simulate the $M_z(\varphi_H)$ of Co layer which interacts with $VO_x$ layer with tilting angle ($\theta_t = -10°$) in Figure 6c. When the external magnetic fields of 0.3 T, 0.4 T, 0.5 T, 2.0 T, are applied we successfully reproduced the experimentally observed phase shift $R_{xy}$ ($\varphi_H$) of Fig. 5b under small in-plane fields. However, this phase shift diminishes when subjected to sufficiently strong fields. It is important to note that we varied the field strength while maintaining a fixed coupling energy between Co and $VO_x$ layers (as depicted in Fig. 6c), which is physically equivalent to applying a fixed field with varying coupling strength (as shown in Fig. 5b II-IV).

Figure 6d illustrates the magnetization configurations generated by the interaction between $VO_x$ and the Co layers in the bilayer at $H_x$ = 0.4 T. The sizeable exchange coupling between $VO_x$ and Co layers makes complex magnetization configurations of Co layer, and it is source of the experimentally observed phase shift. Even though only a few % of Co spins are pinned by ferromagnetic $VO_x$ phase, it affects overall hysteresis and magnetization configurations. We can confirm that the pinned Co spins are just above the ferromagnetic phases of $VO_x$ layer. This indicates that the spin arrangement of the FM layer resulting from the interaction with $VO_x$, and the spin configuration of the FM layer can be controlled by the variation in the ratio of $V^{3+}$ and $V^{4+}$ (or oxidation processes), leading to the emergence of ferromagnetism.

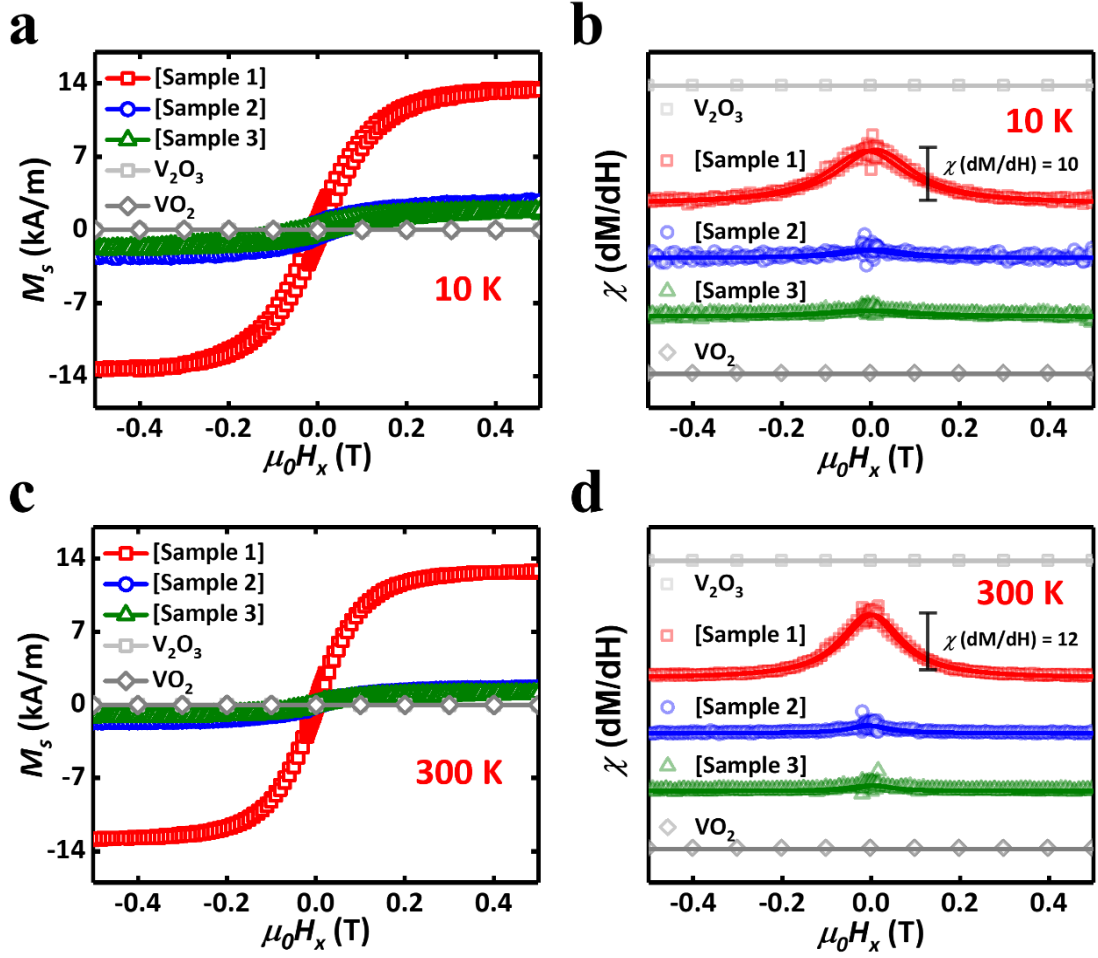

**Figure 1.** The hysteresis loops and magnetic susceptibilities are analyzed to confirm the presence of ferromagnetism in VO$_x$. Figures (a) and (c) present the hysteresis loops measured at 10 and 300 K, respectively. These loops show that VO$_x$ with the lowest oxidation state (Sample 1) exhibits finite magnetization, which is indicative of ferromagnetic behavior. The presence of a hysteresis loop with non-zero coercivity and remanence is a classic signature of ferromagnetism. Figures (b) and (d) display the susceptibility results for each sample. The magnetic susceptibility ($\chi \sim 10$) are significant and cannot be explained by paramagnetic or antiferromagnetic phases alone. This elevated susceptibility further supports the presence of ferromagnetism in the samples. For a more comprehensive understanding, the hysteresis loops and susceptibilities of single crystal V$_2$O$_3$ and VO$_2$ were also examined. The comparative analysis indicates that as the oxidation state of VO$_x$ increases, the magnetic moment associated with ferromagnetism gradually decreases. This trend suggests that ferromagnetism is most pronounced at a specific oxidation level, likely corresponding to a mixed or lower oxidation state. The specifics of Samples 1, 2, and 3, which are detailed in Table I, likely include variations in the oxidation states or structural properties of the VO$_x$ films. These variations help in understanding how different oxidation states or structural configurations influence the magnetic properties.

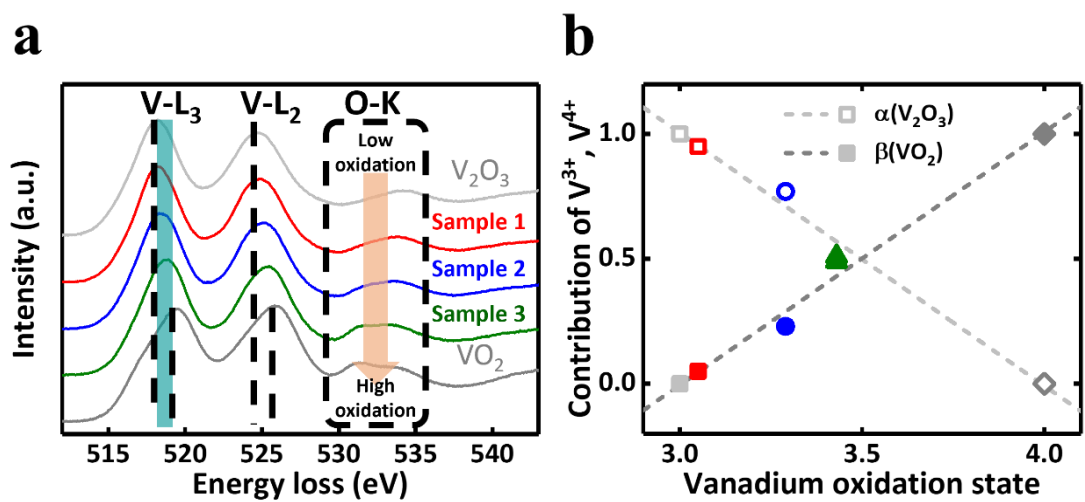

**Figure 2**. In this analysis, transmission electron microscopy-electron energy loss spectroscopy (TEM-EELS) measurements were used to investigate the oxidation states of VO$_x$ samples. The focus was on understanding how the valence states change across different samples and how these relate to known vanadium oxide phases. (a) The TEM-EELS data, which highlights the differences in oxidation states among the VO$_x$ samples. By comparing the data with single crystals of V$_2$O$_3$ (gray, representing V$^{3+}$) and VO$_2$ (gray, representing V$^{4+}$), it is evident that as the oxidation state increases, the material transitions from a V$_2$O$_3$-like state to a VO$_2$-like state. This is particularly noticeable when examining the L$_3$ energy loss, where shifts in energy loss peaks correspond to changes in oxidation states. The appearance of a bump in the O-K edge as the oxidation state increases suggests the coexistence of V$^{3+}$ and V$^{4+}$ within the VO$_x$ samples. This feature is indicative of mixed valence states, which is a common characteristic in transition metal oxides with varying oxidation levels. (b) A quantitative analysis of valence changes using Gaussian functions applied to the L$_3$ edge energy loss variations. The area ratios measured in panel (a) are used to determine the relative proportions of V$^{3+}$ and V$^{4+}$ in each sample. The red, blue, and olive lines correspond to Samples 1, 2, and 3, respectively. The ratio of V$^{3+}$ to V$^{4+}$ for Sample 1, 2, and 3 are 0.95:0.05, 0.77:0.23, and 0.51:0.49, respectively. The results demonstrate a clear trend where the valence state of vanadium in VO$_x$ shifts from predominantly V$^{3+}$ to a mix with increasing V$^{4+}$ content as the oxidation state increases. This progression is crucial for understanding the electronic and magnetic properties of VO$_x$, as the valence state directly influences these characteristics. The ability to tune the V$^{3+}$ to V$^{4+}$ ratio provides a pathway for tailoring the properties of VO$_x$ for specific applications.

| Oxygen flow [sccm] | Area ratio in EELS spectra (Method 1) | $I(L_3)/I(L_2)$ ratio (Method 2) | Energy loss (V-$L_3$) [eV] (Method 3) | Vanadium oxidation state ($V_2O_{3+p}$) |
|---|---|---|---|---|
| $V_2O_3$ | 1:0 | 1.104 | 518.24 | $p = 0$ ($V_2O_3$) |
| 2.5 | 0.95:0.05 | 1.097 | 518.30 | $p = 0.05$ (Sample 1.) |
| 3.5 | 0.77:0.23 | 1.069 | 518.50 | $p = 0.23$ (Sample 2.) |
| 4.5 | 0.51:0.49 | 1.052 | 518.79 | $p = 0.49$ (Sample 3.) |
| $VO_2$ | 0:1 | 0.983 | 519.37 | $p = 1$ ($VO_2$) |

**Table 1.** In this study, experimental measurements were performed to investigate how the oxidation state of VOx affects the contribution of vanadium ions, specifically focusing on the changes in valence states as oxygen flow increases. Three complementary methods were employed using TEM-EELS data to confirm these valence changes. **1) Area Ratio Calculation (Method1)**: The first method involved calculating the ratio of the areas under the EELS spectra corresponding to different oxidation states. This approach provides an estimate of the relative abundance of $V^{3+}$ and $V^{4+}$ ions in the samples. An increase in the area associated with $V^{4+}$ as oxygen flow increases indicates a higher contribution of $V^{4+}$. **2) $I(L_3)/I(L_2)$ Edge Intensity Ratio (Method2)**: The second method analyzed the intensity ratio changes between the $L_3$ and $L_2$ edges. This ratio is sensitive to changes in the oxidation state and can be used to differentiate between $V^{3+}$ and $V^{4+}$. An increasing $I(L_3)/I(L_2)$ ratio with higher oxygen flow further confirms the growing contribution of $V^{4+}$. **3) Energy Loss at the $L_3$ Edge (Method3)**: The third method focused on the changes in energy loss at the $L_3$ edge. By comparing shifts in the $L_3$ edge energy loss, the method provides insights into the oxidation state of vanadium. An increase in energy loss corresponding to $V^{4+}$ suggests an increase in its contribution with higher oxidation levels. These methods collectively confirmed that the contribution of $V^{4+}$ ions increase as the oxidation state of $VO_x$ rises due to higher oxygen flow. To quantitatively represent this change, the notation $V_2O_{3+p}$ was used, where $p$ denotes the relative proportion of $V^{4+}$. This notation effectively captures the gradual transition from a $V_2O_3$-like state towards a more oxidized state with increased $V^{4+}$ content, providing a clear framework for understanding the impact of oxidation on the valence state of vanadium in $VO_x$.

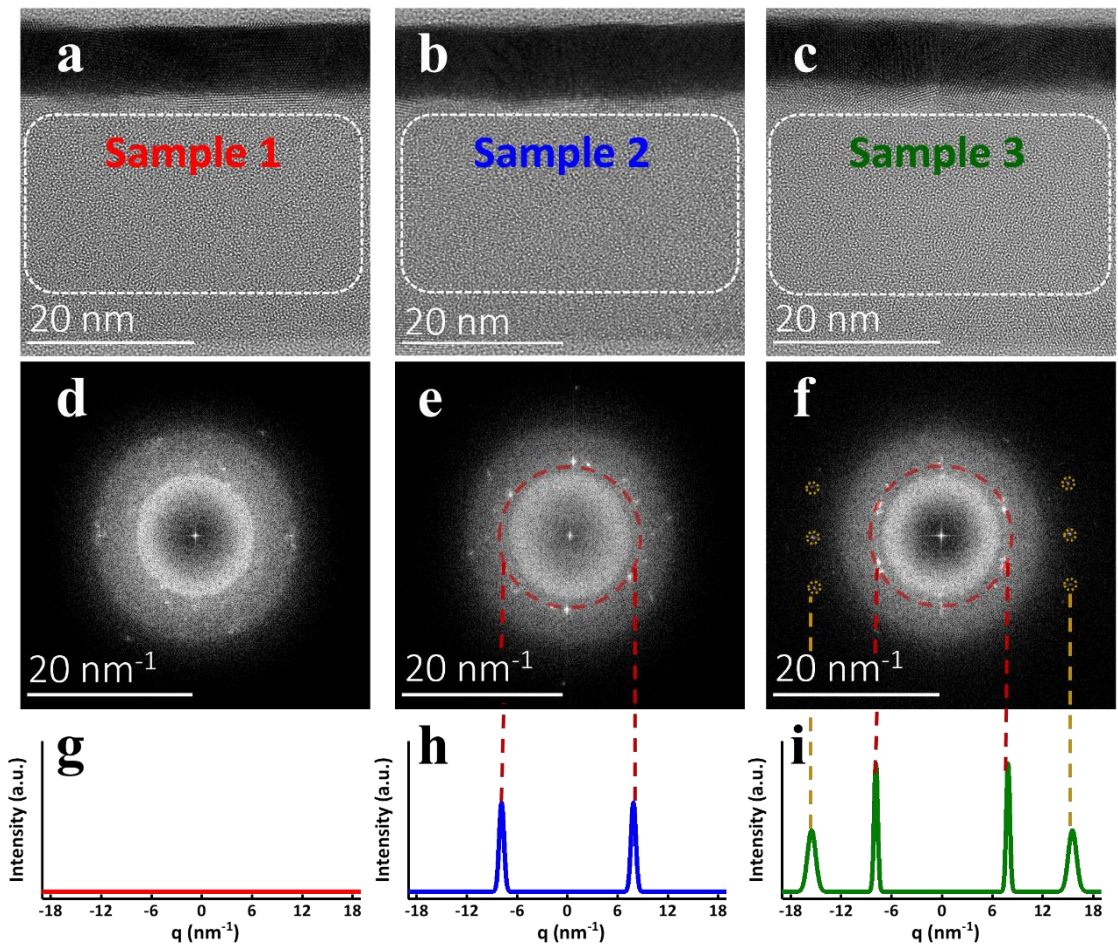

**Figure 3.** The structural analysis of VO$_x$ (Samples 1, 2, 3) with varying oxidation states was conducted using electron diffraction patterns (EDP) and high-resolution transmission electron microscopy (HR-TEM). (a-c) The HR-TEM images of Sample 1, 2, 3. The white dashed boxes highlight the regions corresponding to VO$_x$. (d-f) the EDP obtained by performing fast Fourier transform (FFT) on the HR-TEM images. In (d), the absence of bright spots indicates that Sample 1 has an amorphous structure, suggesting a low oxidation state with minimal lattice ordering. In (e) and (f), an increasing number of bright spots appear as the oxidation state increases. This indicates the formation of a crystalline lattice structure in Samples 2 and 3, respectively. (g-i) the line profiles extracted by integrating the EDP results, focusing on the main rings containing the diffraction spots. These profiles provide quantitative insights into structural changes. In (g), the line profile for Sample 1 shows broad, diffuse peaks, consistent with an amorphous structure and a low oxidation state where V$^{4+}$ ions are sparse. In (h) and (i),

the line profiles for Samples 2 and 3 reveal sharper peaks, indicating increased crystallinity as the oxidation state rises. The analysis suggests that when $V^{4+}$ ions are present in very small amounts, they form isolated domains within an amorphous matrix, leading to the observed ferromagnetism in Sample 1. As the oxidation state increases, the $V^{4+}$ ions begin to organize into a more crystalline structure, characteristic of $VO_2$, which is associated with paramagnetism. This structural transition explains the disappearance of ferromagnetism in Samples 2 and 3 as they become more crystalline and transition towards the properties of $VO_2$.

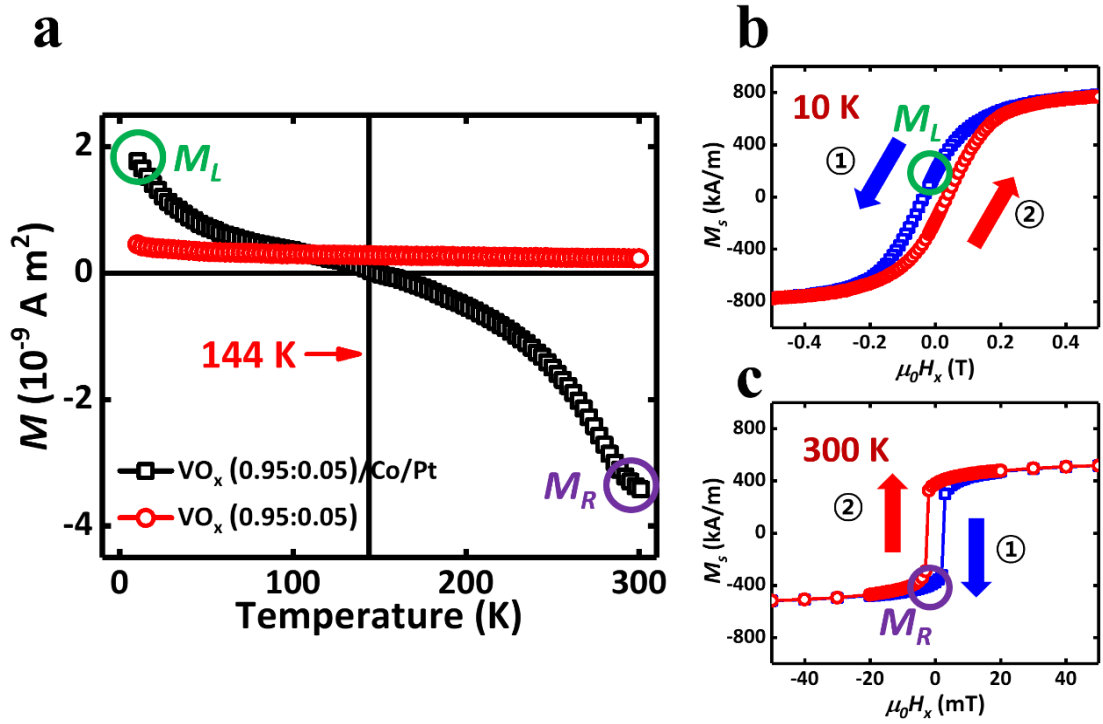

**Figure 4.** The magnetic interaction in the VO$_x$/Co/Pt trilayer system is examined through the $M(T)$ and $M(H)$ curves. (a) The magnetic moment measurement of the VO$_x$ single layer shows a positive magnetic moment across all temperatures. This indicates that the VO$_x$ layer itself exhibits ferromagnetic behavior or a net magnetic moment due to its intrinsic properties, possibly due to isolated V$^{4+}$ ions forming ferromagnetic domains. In the VO$_x$/Co/Pt trilayer, the magnetic moment behavior changes significantly. When the temperature is lowered to 10 K, an external magnetic field is applied, and then removed, the trilayer exhibits a negative magnetic moment above 144 K. The magnetic moment of the trilayer at 10 and 300 K are denoted by "$M_L$" (green circle, low temperature) and "$M_R$" (purple circle, room temperature), and they are shown in (b) and (c) to explain the meaning of negative magnetic moment, respectively. (b) The $M(H)$ curve at 10 K displays a normal hysteresis loop, typical of ferromagnetic materials, indicating that at low temperatures, the trilayer system behaves similarly to a ferromagnet. (c) At room temperature (300 K), the $M(H)$ curve is inverted, suggesting a significant change in the magnetic interaction. This inversion is indicative of antiferromagnetic coupling between the layers, where the magnetic moments of VO$_x$ and Co are aligned antiparallel, leading to a net reduction in the observed magnetic moment. The presence of a negative magnetic moment and the inverted $M(H)$ curve at 300 K suggest that the magnetic moments of VO$_x$ and Co oppose each other, resulting in a net decrease in magnetization.

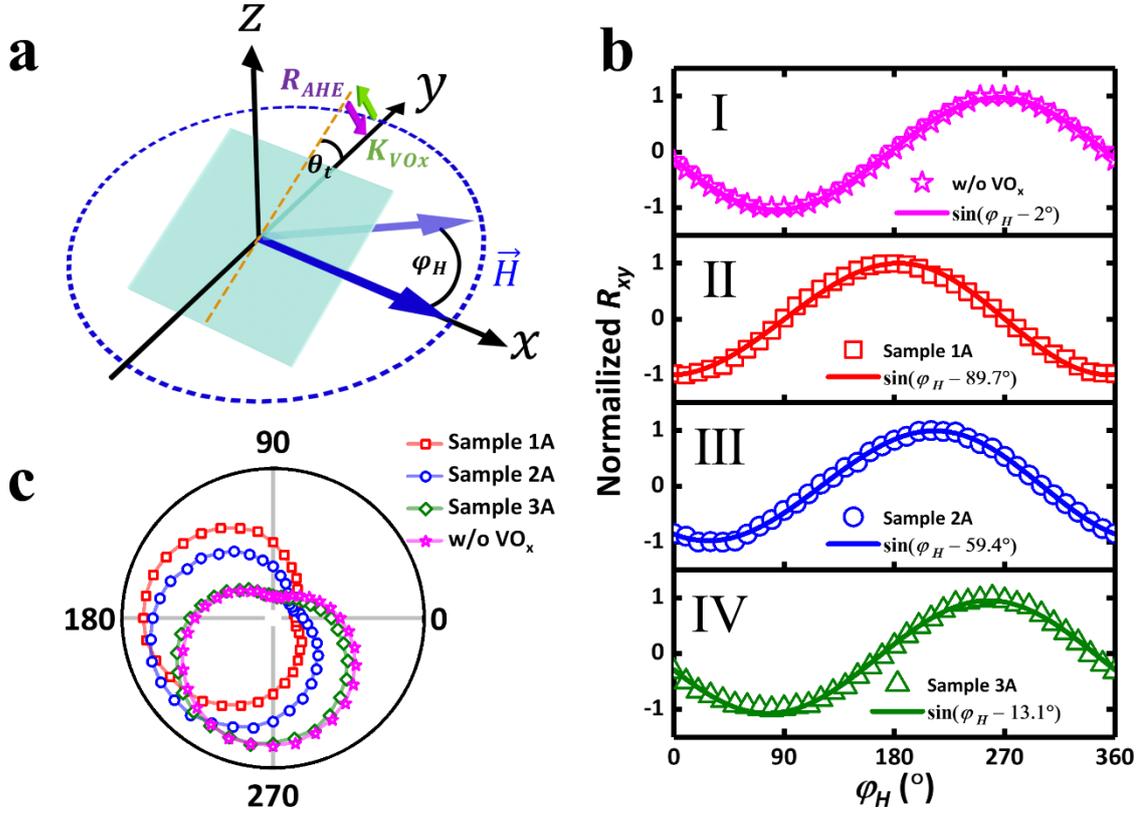

**Figure 5.** The study of the anomalous Hall effect (AHE) in the $VO_x$/Co/Pt trilayer system provides insights into the changes in Co spin arrangement due to the interaction with $VO_x$. (a) The Schematic of experimental setup for measuring the angular-dependent AHE is illustrated. The setup involves applying a fixed angle, $\theta_t$, from y-axis, while varying the rotating azimuthal angle $\varphi_H$ to measure the AHE. This configuration allows for the investigation of how the Co spin arrangement and magnetic interactions change. (b) The pink curve represents the expected behavior of the AHH based on the intrinsic properties of the Co/Pt layer, without considering the interaction with $VO_x$. The red, blue, and green curves are the results of AHE of Sample 1A, 2A, and 3A, respectively. Those three curves show noticeable shift from the simple sine curve. The degrees of the shift are denoted in the figures, and they indicate variations in the Co spin arrangement due to the interaction with $VO_x$.

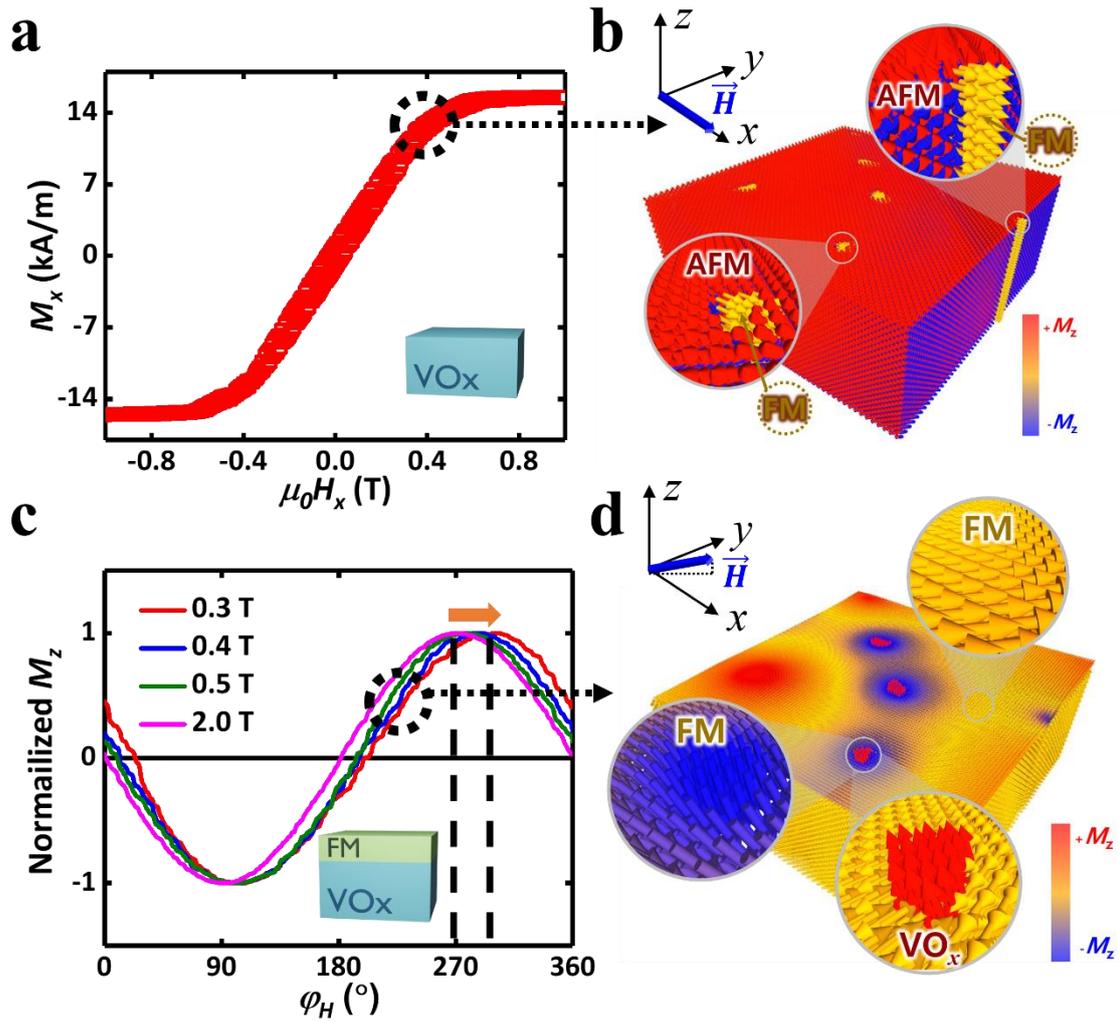

**Figure 6.** The results of micromagnetic simulations for VO$_x$ single layer (a,c) and VO$_x$/Co bilayer (b,d). (a) The hysteresis loop of the *x*-component magnetization as a function of the external in-plane magnetic fields. The saturation magnetization of 15 kA/m is similar to the experimental results (Fig. 1a), contrary to the input value of $M_s(= 1.0 \times 10^6$ A/m) for VO$_x$. (b) Illustrations of the spin configurations at $H_x$ = 0.4 T. The insets show the details of the spin configurations. The spins of the small portion of the ferromagnetic phases are aligned to the field direction, while remaining spins of the antiferromagnetic phases are anti-parallel configurations. The strong pinning between the ferromagnetic and antiferromagnetic phases caused the slanted hysteresis loop. (c) The normalized $M_z$ (corresponding experimental AHE signal) as a function of the in-plane field angle $\varphi_H$ with the external magnetic field is 0.3 T, 0.4 T, 0.5 T, 2.0 T and tilting angle ($\theta_t = -10°$). Experimentally observed phase shift (Fig. 5b)

are successfully reproduced under small in-plane fields (corresponding to the strong coupling strength). And this phase shift diminishes when subjected to sufficiently strong fields (weak coupling). (d) Illustration of the spin configurations of the VO$_x$/Co bilayer at $H_x$ = 0.4 T. Due to the strong coupling between VO$_x$ and Co layers, small portions of Co spins are strongly pinned by the ferromagnetic VO$_x$ phases as shown in the insets. And the coupling causes the phase shift in (c).

## 2. Conclusions

In conclusion, this study identifies the specific changes in magnetic properties that occur at various oxidation states in VO$_x$ thin films, as well as the interactions that arise with the FM layer. Notably, the oxidation of VO$_x$ was tailored to produce different proportions of V$^{3+}$ and V$^{4+}$. This modulation of VO$_x$ composition allows for the adjustment of the ratio between antiferromagnetic and ferromagnetic phases. Furthermore, it indicates the potential to control the interaction with the ferromagnetic layer through variations in oxidation process. To validate these characteristics, anomalous Hall effect measurements were conducted in the $\varphi_H$ direction, with the $z$-direction angle ($\theta_t$) fixed. The results revealed a shift in angle dependence, demonstrating the ability to regulate interactions with the ferromagnetic layer by modifying the oxidation of VO$_x$. Additionally, employing gating with bias voltage to alter the oxidation of VO$_x$ suggests the potential to influence spin-orbit coupling with the resulting ferromagnetic layer. This opens up the possibility of exclusively controlling voltage-induced phenomena, including voltage-controlled magnetic anisotropy, Dzyaloshinskii-Moriya interaction, exchange bias, and domain wall motion, which are essential building blocks of the modern spintronics.

## 3. Experimental section

*Sample preparation*: In this study, VO$_x$, Co, and Pt layers were sequentially deposited on a thermally oxidized silicon wafer using a magnetron sputtering system. The process began with the deposition of a 20 nm thick VO$_x$ layer, achieved through reactive sputtering with a metallic vanadium target. During this deposition, the chamber's working pressure was maintained at 2.0 mTorr. Argon (Ar) and oxygen (O$_2$) gases were introduced into the chamber, with their flow rates precisely controlled by mass flow controllers. The O₂ gas content was varied from 6.9% to 18.4% to explore different oxidation conditions. Following the VO$_x$ layer deposition, a Co layer with a thickness of 1.4 nm was deposited to investigate its magnetic properties. Finally, a 5 nm thick Pt layer was deposited on top of the Co layer.

*Sample Characterization*: In this study, several characterization techniques were employed to investigate the properties of VO$_x$, Co, and Pt films: **1) Binding Energy Spectra**: High-resolution transmission electron microscopy (HR-TEM) was used at room temperature to obtain binding energy spectra of single-layer VO$_x$ films with a thickness of 20 nm. This technique helps in understanding the electronic structure and chemical state of the VO$_x$ layer. **2) Magnetization Measurements**: The magnetic properties were assessed using a Magnetic Properties Measurement System (MPMS) equipped with a superconducting magnet capable of reaching up to 5 T. The magnetic moment as a function of temperature was recorded starting from a saturated state at 10 K under a magnetic field of 1 T. The measurements were then continued up to 300 K after removing the external magnetic field to observe the thermal stability and magnetic behavior of the films. **3) Longitudinal Resistivity**: The angular dependence of longitudinal resistivity was measured at 10 K under a 0.1 T magnetic field using a Physical Properties Measurement System (PPMS). A four-point resistance measurement setup was employed to ensure accurate resistivity measurements by minimizing contact resistance effects. **4) Hall Effect Measurements**: The Hall effect in the VO$_x$/Co/Pt trilayer was investigated using the Van der Pauw method. This method is effective for measuring the Hall voltage and determining the carrier concentration and mobility in thin films. **5) Spin Arrangement Analysis**: To analyze the influence of the VO$_x$ layer on the cobalt layer, spin arrangements were examined by tilting the samples at an angle of $\theta_t = -10°$. This approach helps evaluate the impact of the VO$_x$ layer on the magnetic anisotropy and spin configuration of the Co layer. These comprehensive measurements provide insights into the electronic, magnetic, and transport properties of the VO$_x$/Co/Pt multilayer structure, allowing for a better understanding of the interactions between the layers and the overall behavior of the system.

*Micromagnetic Simulations*: Micromagnetic simulations were conducted using MuMax3[27] to gain a deeper understanding of the interactions between the VO$_x$ layer and the ferromagnetic (FM) Co layer. These simulations numerically solve the Landau-Lifshitz-Gilbert (LLG) equation to model the magnetic behavior of the system. The simulations considered a film with dimensions of $100 \times 100 \times N_z$ nm³, where $N_z$ represents the total number of cells to z-direction. The cell size used in the simulations was $1 \times 1 \times 0.25$ nm³. We set the thickness of VO$_x$ layer is 10 nm and Co layer is 0.5 nm, respectively. For the Co layer, the following material parameters were used: $M_s = 1.0 \times 10^6$ A/m, the exchange stiffness $A_{ex} = 2.0 \times 10^{-11}$ J/m, magnetic anisotropy energy $K = 0.7 \times 10^6$ J/m³, interfacial DM interaction energy density $D = 1.0$ mJ/m². For the VO$_x$ layer, which is assumed to be in a state close to amorphous and thus difficult to define precisely, the following parameters were set: $M_s = 1.0 \times 10^6$ A/m, $A_{ex} = 2.0 \times 10^{-11}$ J/m, $K_{VOx} = 0.05 \times 10^6$ J/m³, $D_{VOx} = 0.001$ mJ/m².

These parameters for VO$_x$ were varied over a wide range to ensure the simulations could reproduce the experimental observations. Although the chosen parameter set successfully replicated the experimental results, it is important to note that these values may not represent the actual material properties of VO$_x$. Instead, they are one possible set of parameters that align with the observed behavior in the experiments. The simulations provide valuable insights into magnetic interactions and help to interpret the experimental findings, although further experimental validation would be necessary to confirm the exact material properties of the VO$_x$ layer.

## Supporting Information

Supporting Information is available from the Wiley Online Library or from the authors.

## Acknowledgements

This work is supported by the National Research Foundation of Korea (NRF) (Grant No. 2021M3F3A2A01037525).

## Conflict of Interest

The authors declare no conflict of interest.

## Data Availability Statement

The data that support the findings of this study are available from the corresponding author upon reasonable request.

# Supporting information

## Control of ferromagnetism of Vanadium Oxide thin films by oxidation states

*Kwonjin Park,*[1] *Jaeyong Cho,*[1] *Soobeom Lee,*[2] *Jaehun Cho,*[3] *Jae-Hyun Ha,*[1] *Jinyong Jung,*[1] *Dongryul Kim,*[1] *Won-Chang Choi,*[1] *Jung-Il Hong,*[1] *and Chun-Yeol You*[1*]

[1]*Department of Emerging Materials Science, Daegu Gyeongbuk Institute of Science & Technology (DGIST), Daegu 42988, Republic of Korea*

[2]*Department of Electrical and Computer Engineering, Shinshu University, Nagano 380-0928, Japan*

[3]*Division of Nanotechnology, Daegu Gyeongbuk Institute of Science & Technology (DGIST), Daegu 42988, Republic of Korea*

*cyyou@dgist.ac.kr

**HR-TEM image of single crystal V₂O₃ and VO₂**

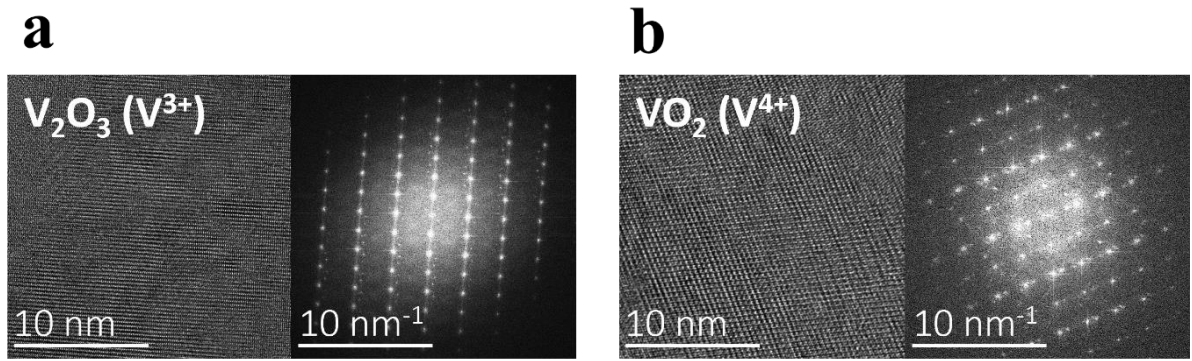

Figure S1. (**a**), (**b**) High resolution transmission electron microscopy (HR-TEM) image and fast Fourier transformation (FFT) showing the single crystals of $V_2O_3$ and $VO_2$ deposited on the Sapphire (0001) substrates. This result determines the vanadium valence states of Sample 1, 2, and 3, and assists to demonstrate the presence of $V_2O_3$ and $VO_2$, composed solely of $V^{3+}$ or $V^{4+}$ ions.

**Methodologies for determining the $V^{3+}$ and $V^{4+}$ states according to the oxidation of $VO_x$ using TEM-EELS**

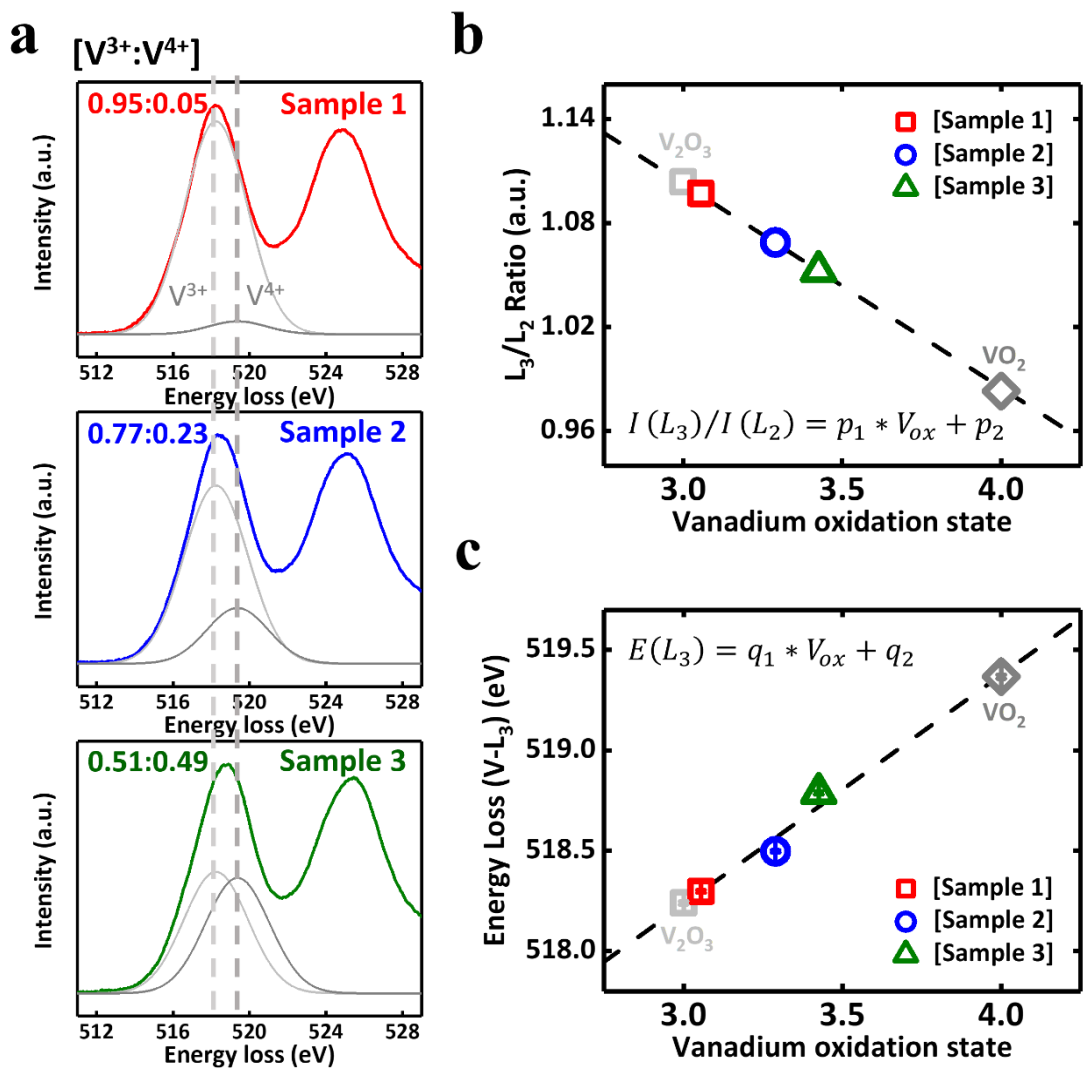

Figure S2. **(a)** Method1: The area ratios of EELS spectra. The results showed area ratios of 0.95:0.05, 0.77:0.23, and 0.51:0.49 for each $VO_x$. **(b)** Method2: The intensity ratios of $L_3$ to $L_2$ of sample 1, 2, 3 and $V_2O_3$ and $VO_2$. **(c)** Method3: The energy loss peak position of the L3 edge of sample 1, 2, 3 and $V_2O_3$ and $VO_2$.

**1) Area Ratio Calculation (Method1)**: For deep analysis of EELS data, we focus EELS spectra on the narrower energy loss range of 512-529 eV in Figure 2 for Sample 1, 2, 3. Consequently, it is expected to have mixed EELS spectra composed of $V^{3+}$ and $V^{4+}$. To clearly distinguish their contributions, single crystals of $V_2O_3$ and $VO_2$ were fabricated for comparative analysis. These are composed with their individual $V^{3+}$ and $V^{4+}$ ions, allowing their Gaussian peaks and energy loss positions to be defined. Initially, the contribution of the oxidation states of VO$_x$ can be explained with the use of the area ratio based on Gaussian functions. In figure S1a, the Gaussian spectrum areas of single crystal V$_2$O$_3$ and VO$_2$ EELS spectra were computed to define the contributions of vanadium ions in VO$_x$. The V$_2$O$_3$ peak originates at 518.24 eV. Additionally, the VO$_2$ peak at 519.37 eV. Using these peaks, multi-Gaussian fitting can be performed for Sample 1, 2, 3. Based on the multi-Gaussian fitting results, contributions of V$^{3+}$ and V$^{4+}$ are determined according to their area ratios. As a result, lowest oxidation (Sample 1) is 0.95:0.05, while more oxidized (Sample 2, 3) are 0.77:0.23 and 0.51:0.49. This indicates the presence of a mixture of V$^{3+}$ and V$^{4+}$ions, and it can be observed that as oxidation increases, the contribution of V$^{4+}$ ($\beta$) increases and V$^{3+}$ ($\alpha$) decreases as shown in Fig. 2b.

**2) I(L$_3$)/I(L$_2$) Edge Intensity Ratio (Method2)**: We analyzed the intensity ratio changes between the L$_3$ and L$_2$ edges of Sample 1, 2, 3. It has been reported that the ratio decreases linearly as oxidation increases in single crystal VO$_x$[1]. The data points for VO$_x$ utilized with the linear relation, $I(L_3)/I(L_2) = p_1 * V_{ox} + p_2$, where $V_{ox}$ represents the vanadium oxidation state, and $p_1$ and $p_2$ obtained values of -0.12 and 1.46 from V$_2$O$_3$ and VO$_2$, respectively. From these results, VO$_x$ ($V_{ox}$) can be defined from the $I(L_3)/I(L_2)$ values of Sample 1, 2, 3. As we mentioned, the results well matched with the Method1 in Table I.

**3) Energy Loss peak of the L$_3$ Edge (Method3):** According to the S. Kalavathi[1], the energy loss peak of L$_3$ edge also reflect the oxidation state, so another linear relation of the energy loss peak of L$_3$ edge is given: $E(L_3) = q_1 * V_{ox} + q_2$, and $q_1$ and $q_2$ are determined from the peak position of V$_2$O$_3$ and VO$_2$. ($q_1 = 1.143, q_2 = 514.8$). We depicted the linear line ($q_1 * V_{ox} + q_2$) with $E(L_3)$ for Sample 1, 2, 3 and corresponding $V_{ox}$ values obtained from Method2 in Figure S1c. Surprisingly enough, data points of Sample 1, 2, 3 are falling into the linear line, and it implies our methodologies for determining the V3+ and V4+ states are consistence. The results are summarized in Table I.

**Note S1: Detail of the azimuthal angle $\varphi_H$ dependent Hall signal experiments.**

To investigate the changes in spin arrangement arising from the interaction between $VO_x$ and Co layers, we measured the Hall effect as a function of the azimuthal angle direction $\varphi_H$ with small tilting geometry as shown in Fig. 5a. Generally assuming an strong enough magnetic external magnetic field, the $\hat{m} = (\sin\theta_M \cos\varphi_M, \sin\theta_M \sin\varphi_M, \cos\theta_M)$ is parallel to the external magnetic field, so that we know the $\vec{M}$. In this case, the total Hall resistance can be defined as a function of polar and azimuthal angles.

$$R_{xy} = \frac{1}{2}\Delta R_{AHE} \cos\theta_M + \frac{1}{2}\Delta R_{PHE} \sin^2\theta_M \sin 2\varphi_M$$

$R_{AHE}$ and $R_{PHE}$ expressed the anomalous Hall effect and planar Hall effect [2]. However, in our experiment, we used tilted geometry to investigate the coupling between $VO_x$ and Co layers. To incorporate the parameter for tilting in the $\theta_t$ direction, we used Euler equation. The magnetic field rotating only in the x-y plane before tilting is expressed as $\vec{H} = H_{ext}(\cos\varphi_H, \sin\varphi_H, 0)$. Using the Euler equation to depict the rotation of the axis, the calculation with respect to the x-axis, defined as

$$\begin{bmatrix} 1 & 0 & 0 \\ 0 & \cos\theta_t & -\sin\theta_t \\ 0 & \sin\theta_t & \cos\theta_t \end{bmatrix} \begin{bmatrix} \cos\varphi_H \\ \sin\varphi_H \\ 0 \end{bmatrix} = \begin{bmatrix} \cos\varphi_H \\ \cos\theta_t \sin\varphi_H \\ \sin\theta_t \sin\varphi_H \end{bmatrix}$$

Thus, the applied external field in our tilted geometry rewritten $\widehat{H_{tilted}} = (\cos\varphi_H, \cos\theta_t \sin\varphi_H, \sin\theta_t \sin\varphi_H)$. And assuming a sufficiently large external magnetic field, and the direction of the magnetization are parallel to the external magnetic field ($\hat{m} \parallel \widehat{H_{tilted}}$), we can rewrite

$$(\sin\theta_M \cos\varphi_M, \sin\theta_M \sin\varphi_M, \cos\theta_M) = (\cos\varphi_H, \cos\theta_t \sin\varphi_H, \sin\theta_t \sin\varphi_H)$$

By using this, when calculating the total Hall resistance in the tilted geometry with $\theta_t$,

$$\cos\theta_M = \sin\theta_t \sin\varphi_H$$

$$\sin\theta_M = \sqrt{1 - \sin^2\theta_t \sin^2\varphi_H}$$

$$\cos\varphi_M = \frac{\cos\varphi_H}{\sqrt{1 - \sin^2\theta_t \sin^2\varphi_H}}$$

$$\sin\varphi_M = \frac{\cos\theta_t \sin\varphi_H}{\sqrt{1 - \sin^2\theta_t \sin^2\varphi_H}}$$

$$R_{xy} = \frac{1}{2}\Delta R_{AHE}(\sin\theta_t \sin\varphi_H) + \frac{1}{2}\Delta R_{PHE} \cos\theta_t \sin 2\varphi_H \qquad (S1)$$

This result is important in our tilted experimental geometry. Here, $\theta_t$ is fixed value in the experiment, it becomes a constant value. Eq. S1 shows $\sin\varphi_H$ ($\sin 2\varphi_H$) dependence for AHE (PHE). It must be mentioned that there is no azimuthal angle dependence ($\sin\varphi_H$) in standard AHE experiment, but we have $\sin\varphi_H$ dependent signals for AHE experiments (Fig. 5b-c) in order to detect in-plane phase shift, which reflect the coupling between Co and $VO_x$ layers. Even though we applied not strong enough external field, the magnetization never aligned to the external magnetic field, however, it is enough to generate $\sin\varphi_H$ dependent. Thus, the shift due to the interaction between $VO_x$ and Co can be expressed as $\sin(\varphi_H + x)$. For $VO_x$ with the lowest oxidation, this implies a change in spin arrangement due to antiferro-coupling. In cases of higher oxidation in $VO_x$, antiferro-coupling weakens, gradually like the general case of $\sin\varphi_H$.

**Hall signal with the varying external magnetic field strength**

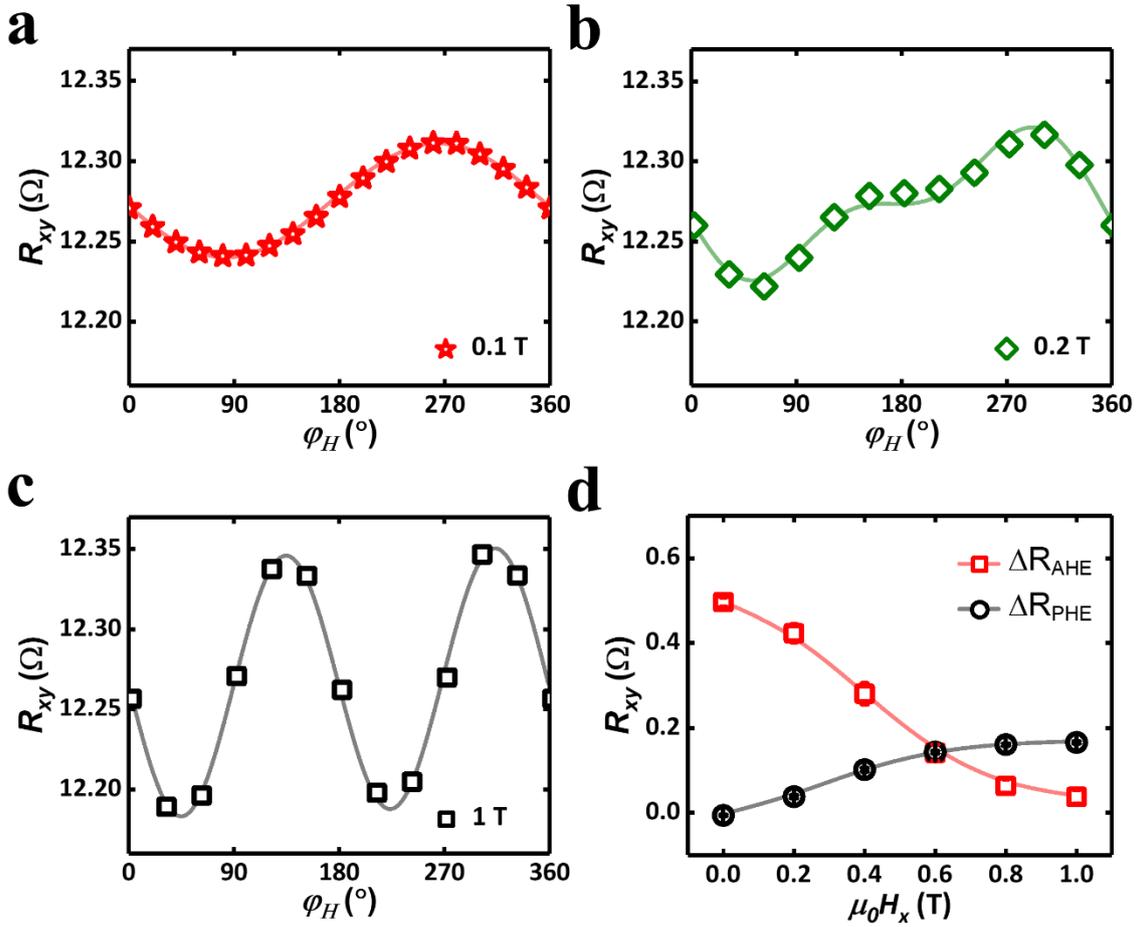

Figure S3. Trend of $R_{xy}$ with respect to the azimuthal $\varphi_H$ angle of external magnetic field (**a**) 0.1 T, (**b**) 0.2 T, and (**c**) 1T. (**d**) represents the contributions of AHE and PHE components with the external magnetic field strength.

The measured Hall signals ($R_{xy}$) of the control sample (Co/Pt) are sum of the AHE and PHE contributions. In Eq. (1), we assume sufficiently large external field so that the magnetization of sample aligned to the external magnetic field ($\theta_M = \frac{\pi}{2} - \theta_t$). However, in our experimental situation, we applied moderate strength field ($\theta_M \neq \frac{\pi}{2} - \theta_t$), so that $R_{xy}$ signals are mixture of AHE and PHE. Figure S3a-c applies external magnetic fields of 0.1 T, 0.2 T, 1T, respectively. In Figure S3a, the AHE dominates due to the insufficient external magnetic field, resulting in the appearance of the $M_z$ component $\sin \varphi_H$. Additionally, in figure S3b, the components of AHE and the planar Hall effect (PHE) are similar, leading to the trend of $\sin \varphi_H + \sin 2\varphi_M$. In

Figure S3c, a sufficiently strong external magnetic field causes the PHE to dominate resulting in the trend of $\sin 2\varphi_H$. Figure S3d quantitatively expresses the changes in AHE and PHE with the magnitude of the external magnetic field. Similarly, it can be observed that when the external field is small, the AHE term is relatively larger than the PHE term, whereas when it is large, the PHE term is relatively larger. Ultimately, by varying the magnitude of the external magnetic field, this demonstrates changes in the trend resulting from the different sizes of AHE and PHE. In the main text, we carried out the experiment with 0.1 T field, so that the AHE signal is dominant, but still poses the information of the in-plane component dependent signal.